\documentclass{amsart}
\usepackage{amssymb}
\usepackage{color}
\usepackage{a4}
\begin{document}
\newtheorem{theorem}{Theorem}[section]
\newtheorem{lemma}[theorem]{Lemma}
\newtheorem{remark}[theorem]{Remark}
\newtheorem{definition}[theorem]{Definition}
\newtheorem{corollary}[theorem]{Corollary}
\newtheorem{example}[theorem]{Example}
\makeatletter
 \renewcommand{\theequation}{%
  \thesection.\alph{equation}}
 \@addtoreset{equation}{section}
 \makeatother
\def\id{\operatorname{Id}}
\def\qedbox{\hbox{$\rlap{$\sqcap$}\sqcup$}}
\def\ffrac#1#2{{\textstyle\frac{#1}{#2}}}
\def\Tr{\operatorname{Tr}}
\def\nn{\nonumber}
\def\bea{\begin{array}}
\def\eea{\end{array}}
\def\beq{\begin{eqnarray}}
\def\eeq{\end{eqnarray}}
\newcommand{\nats}{\mathbb{N}}
\newcommand{\reals}{\mathbb{R}}
\newcommand{\cal}{\mathcal}
\title[Eta Invariants]
{Eta invariants with spectral boundary conditions}
\author{P. Gilkey, K. Kirsten, and J. H. Park}
\begin{address}{PG: Math. Dept., University of Oregon, Eugene, Or 97403, USA}\end{address}
\begin{email}{gilkey@darkwing.uoregon.edu}\end{email}
\begin{address}{KK: Department of Mathematics, Baylor University \\
Waco, TX 76798, USA}\end{address}
\begin{email}{Klaus\_Kirsten@baylor.edu}\end{email}
\begin{address}{JHP: Dept. of Computer Engineering, Honam  University,
  Gwangju 506-714 Korea}\end{address}\begin{email}{jhpark@honam.ac.kr}\end{email}
\begin{abstract} We study the asymptotics of the heat trace $\Tr\{fPe^{-tP^2}\}$ where $P$ is
an operator of Dirac type, where $f$ is an auxiliary smooth
smearing function which is used to localize the problem, and where
we impose spectral boundary conditions. Using functorial
techniques and special case calculations, the boundary part of the
leading coefficients in the asymptotic expansion is found.
\end{abstract}
\maketitle
\section{Introduction}\label{sect-1}
Let $P$ be an operator of Dirac type with leading symbol $\gamma$ on a vector bundle $V$ over a compact $m$
dimensional Riemannian manifold $M$ with smooth boundary $\partial M$.
One may choose a Hermitian inner product
$(\cdot,\cdot)$ and a Hermitian connection $\nabla$ on $V$ so that $\gamma$ is skew-adjoint and so that
$\nabla\gamma=0$ \cite{BrGi-92}; such structures are said to be {\it compatible} with the given
Clifford module structure
$\gamma$. Let indices
$i,j$ range from $1$ to
$m$ and index a local orthonormal frame $\{e_i\}$ for the tangent bundle of $M$. We adopt the Einstein convention
and sum over repeated indices to expand
$$P=\gamma_i\nabla_{e_i}+\psi_P$$
where $\psi_P$ is a smooth endomorphism of $V$; the sign convention for $\psi_P$ differs from that in
\cite{BrGi-92,BrGi-92a}. Note that the matrices
$\gamma_i$ are skew-adjoint endomorphisms of $V$ satisfying the {\it Clifford commutation relations}
$$\gamma_i\gamma_j+\gamma_j\gamma_i=-2\delta_{ij}\,.$$

If $\partial M$ is non-empty, then we must impose suitable
boundary conditions. For $m$ even, $P$ always admits local
elliptic boundary conditions; see, for example, the discussion of
{\it bag boundary conditions} in \cite{BGKS03,BSW02}. However, if
$m$ is odd, there is a topological obstruction to the existence of
local boundary conditions for certain operators. We therefore
introduce {\it spectral boundary conditions}; these boundary
conditions, which are defined regardless of the parity of $m$,
play a crucial role in the index theorem for manifolds with
boundary \cite{AtPaSi75a}.

Spectral boundary conditions were first introduced by Atiyah et.
al. \cite{AtPaSi75a} in their study of Hirzebruch signature
theorem for manifolds with non-empty boundary. The crucial point
at issue was the definition of a suitable elliptic boundary value
problem for the signature operator whose index was the signature
of the manifold. Although the de Rham complex, whose index is the
Euler characteristic, admits local boundary conditions (i.e.
boundary conditions which are a mixture of Robin and Dirichlet),
the signature complex does not. The signature complex does admit
spectral boundary conditions -- these are pseudo-differential
boundary conditions -- and their introduction was a crucial
turning point.

In order to describe these boundary conditions, near the boundary
we choose a local orthonormal frame so $e_m$ is the inward unit
geodesic normal vector field and $\{e_a\}$ for $1\le a\le m-1$ is
the induced orthonormal frame for the tangent bundle of the
boundary. Let
$$\gamma_a^T:=-\gamma_m\gamma_a$$
be the induced tangential Clifford module structure. Let $\psi_A$ be an auxiliary smooth endomorphism of
$V|_{\partial M}$. Consider the auxiliary operator of Dirac type on $V|_{\partial M}$
$$A:=\gamma_a^T\nabla_{e_a}+\psi_A\,.$$

Assume
$A$ has no purely imaginary eigensections. Let $C$ be a suitable contour in the complex plane
containing the spectrum of $A$ with positive real part. Let
$$\Pi_A^+:=\textstyle\ffrac1{2\pi\sqrt{-1}}\int_C(A-\lambda)^{-1}d\lambda$$
be  the {\it spectral boundary operator}; $\Pi_A^+$ is spectral projection on the generalized eigenspaces
associated to eigenvalues with positive real part. Let
$P_A$ be the realization of $P$ with respect to the boundary conditions defined by $\Pi_A^+$.

The spectral information regarding this boundary value problem is
encoded in the zeta function and the eta function which are
defined as follows. Assume for the sake of simplicity that $P_A$
is self-adjoint (we will be forced to drop this requirement
presently). Let $(\lambda_l , \varphi_l)$ be a spectral resolution
of $P_A$; $\{\varphi_l\}$ is a complete orthonormal basis for
$L^2(V)$ such that $$ \left. P_A \varphi _l = \lambda _l \varphi
_l, \quad \quad \Pi_A^+ \varphi _l \right|_{\partial M} = 0 .$$
Then the zeta function associated with $P_A^2$ is , \beq \zeta (s;
P,A) := \sum_{\lambda_l \neq 0} (\lambda_l ^2) ^{-s}
\label{defzeta}\eeq valid for $\Re s > m/2$. Note, the fact that
$P_A$ as a first order differentiable operator can have positive
and negative eigenvalues does not enter the zeta function of the
Laplace-type operator $P_A^2$. However, the sign is taken into
account defining the eta function of $P_A$, \beq \eta (s; P,A) :=
\sum_{\lambda_l \neq 0} \mbox{sign} (\lambda_l ) |\lambda_l|^{-s}
, \label{defeta}\eeq valid for $\Re s > m-1$. Similarly, one can
define $\zeta (s; A)$ and $\eta (s;A)$; since $\partial M$ is
closed, there is no boundary condition required.

Although the above series representations (\ref{defzeta}) and
(\ref{defeta}) are valid only in the given region of the complex
$s$-plane, the eta and zeta functions can be analytically
continued to meromorphic functions defined on the whole complex
plane. The value $\eta (0;P,A)$ is essential for the description
of the index of $P_A$.

One can also discuss the heat trace. Let $\phi$ be the `initial
temperature distribution' and let $u_\phi (t,x)$ denote the
subsequent temperature distribution. Then $u_\phi (t,x)$ is
determined by the equations
$$\left. (\partial _t + P^2) u_\phi (t,x) = 0, \quad \quad \Pi_A^+ u_\phi
\right|_{\partial M} =0\quad \quad \mbox{and}\quad \quad u_\phi
(0,x) = \phi (x) .$$ The associated fundamental solution ${\cal
K}: \phi \to u_\phi$ is then given by ${\cal K}= e^{-tP_A^2}$. Let
$dx$ and $dy$ be the Riemannian measures on $M$ and on $\partial
M$ respectively. There exists a smooth endomorphism-valued kernel
$K(t,x,\bar x, P^2, A): V_{\bar x} \to V_x$ such that $$u_\phi
(t,x) = ({\cal K} \phi ) (t,x) = \int\limits_M K (t,x,\bar x , P^2
, A) \phi (\bar x) d\bar x.$$ For fixed $t$, the operator ${\cal
K} (t): \phi \to \phi (t,\cdot )$ is of trace class. For $F\in
C^\infty(\operatorname{End}(V))$ a smooth auxiliary smearing
endomorphism used for localizing the problem, we define \beq
a^\zeta ( F, P, A) &:= &\mbox{Tr}_{L^2} \left( F e^{-t
P_A^2}\right) = \int\limits_M \mbox{Tr}_{V_x} \left( F(x)
K(t,x,x,P^2,A) \right) dx \nn\\
a^\eta (F, P, A) &:=& \mbox{Tr}_{L^2} \left( F P_A e^{-t P_A^2}
\right) =\int\limits_M \mbox{Tr}_{V_x} \left( F(x) P_A
K(t,x,x,P^2,A) \right) dx.\nn\eeq

Grubb and Seeley \cite{GrSe95} showed that there are asymptotic
expansions as $t\downarrow 0^+$ of the form:
\begin{equation}\label{eqn-1.a}
\begin{array}{ll}
a^\zeta ( F, P, A) \sim\textstyle\sum_{n=0}^{m-1} a_n^\zeta(F,P,A)t^{(n-m)/2}+{\cal O} (\ln t),\\
a^\eta (F, P, A)\sim\textstyle\sum_{n=0}^{m-1}
a_n^\eta(F,P,A)t^{(n-m-1)/2}\, + {\cal O} (t^{1/2} \ln t).
\vphantom{\vrule height 11pt}\end{array}
\end{equation} We refer to the coefficients $a_n^\zeta$ and $a_n^\eta$
as the zeta and eta invariants respectively.

We note that there are in fact full asymptotic series for
$a^\zeta$ and $a^\eta$. However non-local terms and log terms
arise when $n\ge m$. Since we shall assume that $n<m$, these terms
play no role for us. We shall normally assume that $F=f\cdot\id$
where $f\in C^\infty(M)$ is scalar valued, but it will be
convenient occasionally to have this more general setting
available.

The Mellin transform can be used to relate the zeta and eta
functions and the small-$t$ asymptotic expansion of the
heat-trace. For $f=1$, one has \cite{Gb,seel} \beq \mbox{Res
}\zeta \left( \frac{m-n} 2; P,A\right) &=& \frac{
a_n^{\zeta} (1,P,A) }{\Gamma \left( \frac{ m-n} 2 \right)} ,\nn\\
\mbox{Res }\eta (m-n; P,A) &=& \frac{ 2 a_n^\eta (1,P,A)}{\Gamma
\left( \frac{m-n+1} 2 \right) } .\nn\eeq Similar formulas hold for
general endomorphism $F$; this will play an important role in our
subsequent development.

The heat trace coefficients $a_n^\zeta$ and $a_n^\eta$ of Equation
(\ref{eqn-1.a}) are locally computable for $n<m$; they play a
crucial role in many areas. For example, the particular
coefficient $a_m^\zeta$ is relevant in the quantum mechanics of
closed cosmologies, where it describes how quantum effects modify
the behaviour of the universe near classical singularities
\cite{new2,new2a,new1}. More generally, the leading coefficients
$a_n^\zeta$, $n=0,1,...,m$ are needed in different quantum field
theories. These theories are generically plagued by divergences
which are removed by a renormalization. In the zeta function
scheme \cite{new3}, as well as in the framework of recent
developments of algebraic quantum field theory \cite{moretti}, at
one-loop, divergences are completely described by the leading
coefficients. As a result, their knowledge is equivalent to a
knowledge of the one-loop renormalization group equations
\cite{new4}, which provides one reason for the consideration of
heat kernel coefficients in physics. In addition, if an exact
evaluation of relevant quantities is not possible, asymptotic
expansions are often very useful and most suitably given in terms
of heat kernel coefficients \cite{new5,new6}. In this context of
quantum field theories, apart from cosmology, spectral boundary
conditions most prominently make their appearance in bag models
where they have important advantages over local elliptic boundary
conditions. In particular, it is the only self-adjoint boundary
condition which respects the charge conjugation property and the
so-called $\gamma_5$ symmetry \cite{new7,new8,new9}. In Euclidean
gauge field theories, this property enables one to consider a
compactified Dirac problem where spectral information such as
functional determinants are directly related to the original
problem \cite{new10,new11}.

Whereas the above relates to $a_n^\zeta$, the $\eta$-function
arises in the analysis of fermion number fractionization in
different field theory models \cite{new12,new13,new14}. The
fermion number $N$ is a transcendental function of the parameters
of the theory and is related to $\eta (0;H,A)$ of the pertinent
Dirac Hamiltonian $H$ and boundary operator $A$. In a simplified
consideration \cite{new15} the fermion number of the vacuum will
be formally obtained by filling the Dirac sea, \beq N &=&
\left[\mbox{number of negative-energy
states of $H$}\right]\nn\\
&=& \frac 1 2 \left\{\left[\left(\mbox{numb. of pos.-en. states of
H}\right)+\left(\mbox{numb. of neg.-en. states of
H}\right)\right] \right.\nn\\
& &\left.-\left[\left(\mbox{numb. of pos.-en. states of H}\right)-
\left(\mbox{numb. of neg.-en. states of
H}\right)\right]\right\}\nn\eeq
Regularizing this divergent expression it becomes $(1/2) [ 0- \eta
(0; H,A) ]$. A rigorous proof can be found in \cite{new16}.

Furthermore, interpreting (\ref{defzeta}) and (\ref{defeta}) as a
moment problem for the spectral density function, even and odd
part of the density can be found provided $\zeta (s;H,A)$ and
$\eta (s;H,A)$ can be evaluated \cite{new17}. Knowledge of the
leading coefficients $a_n^\zeta$ respectively $a_n^\eta$ amounts
to an asymptotic knowledge of the even and odd part of the density
for large eigenvalues $|\lambda_l|$ opening up the possibility for
the approximate evaluation of different quantities in quantum
field theories as for example the finite temperature induced
fermion number \cite{new13}.

The invariants $a_n^\zeta$ have been studied extensively
\cite{DoGiKi99,GiKi-03,GrSe95,GrSe-96}; the invariants $a_n^\eta$
have received a bit less attention. We may decompose
\begin{eqnarray*}
&&a_n^\zeta(F,P,A)=a_n^{\zeta,M}(F,P)+a_n^{\zeta,\partial M}(F,P,A),\quad\text{and}\\
&&a_n^\eta(F,P,A)=a_n^{\eta,M}(F,P)+a_n^{\eta,\partial M}(F,P,A)
\end{eqnarray*}
as the sum of interior and boundary contributions. There exist local endomorphism valued invariants
$e_n^{\zeta,M}(x,P)$ and $e_n^{\eta,M}(x,P)$, which are homogeneous of weight $n$ in the jets of the total
symbol of
$P$, so that
\begin{eqnarray*}
&&\textstyle a_n^{\zeta,M}(F,P)=\int_M\Tr\{F(x) e_n^{\zeta,M}(x,P)\}dx,\quad\text{and}\\
&&\textstyle a_n^{\eta,M}(F,P)=\int_M\Tr\{F(x)e_n^{\eta,M}(x,P)\}dx\,.\end{eqnarray*}
We note there is a
parity constraint for the interior invariants;
$$
a_n^{\zeta,M}=0\quad\text{if}\quad n\quad\text{is odd}\quad\text{and}\quad
a_n^{\eta,M}=0\quad\text{if}\quad n\quad\text{is even}\,.
$$
Formulae for the invariants $a_n^{\zeta,M}$ for
$n=0,2,4,6,8$ follow from work of
\cite{AmBeOc89,Av90,Gi75b,McSi67}; similar formulae for the invariants $a_n^{\eta,M}$ are known for $n=1,3$
\cite{BrGi-92}.

Let $\nabla_m^kF$ denote the $k^{\operatorname{th}}$ normal covariant derivative of the
endomorphism $F$. There are local invariants $e_{n,k}^{\zeta,\partial M}(y,P,A)$ and
$e_{n,k}^{\eta,\partial M}(y,P,A)$ which are homogeneous of weight $n-k-1$ in the jets of the total symbol of $P$
and of $A$ so that
\begin{eqnarray*}
&&\textstyle a_n^{\zeta,\partial M}(F,P,A)
   =\textstyle\sum_{k<n}\int_{\partial M}\Tr\{\nabla_m^kF(y)\cdot e_{n,k}^{\zeta,\partial
M}(y,P,A)\}dy\quad\text{and}\\
&&\textstyle a_n^{\eta,\partial M}(F,P,A)
   =\textstyle\sum_{k<n}\int_{\partial M}\Tr\{\nabla_m^kF(y)\cdot
e_{n,k}^{\eta,\partial M}(y,P,A)\}dy\,.\end{eqnarray*}

Let $\Omega_{ij}$ be the curvature of the connection $\nabla$. We define
\begin{eqnarray*}
&&W_{ij}:=\Omega_{ij}-\ffrac14R_{ijkl}\gamma_k\gamma_\ell,\quad
\beta(m):=\Gamma(\textstyle\frac m2)\Gamma(\textstyle\frac12)^{-1}\Gamma(\frac{m+1}2)^{-1},
\quad\text{and}\\
&&E:=\ffrac12(\psi_{P;i}\gamma_i-\gamma_i\psi_{P;i})-\psi_P^2
-\ffrac14(\psi_P\gamma_i+\gamma_i\psi_P)(\psi_P\gamma_i+\gamma_i\psi_P)\\
&&\hphantom{E:}
-\ffrac12\gamma_i\gamma_jW_{ij}-\ffrac14\tau\,.
\end{eqnarray*}
Let $\tau:=R_{ijji}$ be the scalar curvature and let $L_{ab}$ be
the second fundamental form. We can use
\cite{BrGi-92,DoGiKi99,GiKi-03} to see:
\begin{theorem}\label{thm-1.1}
If $P_A$ is self-adjoint, if $A$ is self-adjoint, and if $F=f\cdot\id$ is scalar,\begin{enumerate}
\item $a_0^\zeta(F,P,A) =  (4\pi )^{-m/2} \int_{M}f\Tr\{\id\}$dx.
\smallbreak\item
      If $m\geq 2$, $a_1^\zeta(F,P,A) = (4\pi)^{-(m-1)/2}\frac14(\beta(m)-1)\int_{\partial M}f\Tr\{\id\}dy$.
\smallbreak\item If $m\geq 3$, $a_2^\zeta(F,P,A)
     =(4\pi)^{-m/2}\int_Mf\Tr\{\ffrac16\tau\id+E\}dx$\newline$
   +(4\pi)^{-m/2}\int_{\partial M}\{\frac13(1-
     \frac34\pi \beta(m))L_{aa} f$
     $-\frac{m-1}{2(m-2)}(1-
        \frac12 \pi \beta(m))f_{;m}\}\Tr\{\id\}dy$.
\end{enumerate}\end{theorem}
We refer to \cite{GiKi-03} for the corresponding computation of
$a_3^\zeta(f,D,\mathcal{B})$. In this note, we establish formulas
for $a_n^\eta$ without self-adjointness assumptions:

\begin{theorem}\label{thm-1.2}
Let $F=f\cdot\id$ be scalar, then
 \ \begin{enumerate}\item
$a_0^\eta(F,P,A)=0$. \smallbreak\item If $m\ge2$,
$a_1^\eta(F,P,A)=(4\pi)^{-m/2}(1-m)\int_Mf\Tr\{\psi_P\}dx$.
\smallbreak\item If $m\ge3$,
$a_2^\eta(F,P,A)=(4\pi)^{-(m-1)/2}\int_{\partial
M}f\Tr\{\ffrac{2-m}4(\beta(m)-1)\psi_P$\smallbreak$-\ffrac{1}4
\beta (m) \gamma_m\psi_A\}dy$. \smallbreak\item If $m\ge4$,
$a_3^\eta(F,P,A)=-\ffrac1{12}(4\pi)^{-m/2}\int_Mf\Tr\{[2(m-1)\psi_{P;i}$
  \smallbreak
      $+3(4-m)\psi_P \gamma_i\psi_P
       +3\gamma_j\psi_P\gamma_j\gamma_i\psi_P]_{;i}
      +(3-m) \{\tau \psi_P +6 \gamma_i\gamma_jW_{ij} \psi_P$
   \smallbreak
     $-6\psi_P\psi_{P;i}\gamma_i
      +(4-m)\psi_P \psi_P \psi_P +3 \psi_P \psi_P\gamma_i\psi_P\gamma_i
      \} \big\}dx$
    \smallbreak
     $+(4\pi)^{-m/2}\textstyle\int_{\partial M}\Tr\{
\ffrac{(m-3)(m-1)}{2(m-2)}(1-\ffrac12\pi\beta(m))f_{;m}\psi_P$\smallbreak
$-f\ffrac{(3-m)^2}{4(m-2)}(\psi_P\psi_A+\gamma_m\psi_P\gamma_m\psi_A)
+f\ffrac{3-m}3(1-\ffrac34\pi\beta(m))L_{aa}\psi_P$\smallbreak$
+f\{\ffrac{(m-3)(m-1)}{2(m-2)}(1-\ffrac12\pi\beta(m))-\frac16(m-1)\}\psi_{P;m}$\smallbreak
$-f\ffrac{3-m}{4(m-2)}(\gamma_a^T\psi_P\gamma_a^T\psi_A-\gamma_a\psi_P\gamma_a\psi_A
+2\gamma_m\gamma_a^T\psi_A\gamma_a^T\psi_A)$\smallbreak
$+\ffrac{1}{2(m-2)}\left(1-\ffrac{1}2 \pi (m-1) \beta (m)\right)
(\ffrac{m-3}{1-m}fL_{aa}+f_{;m})\gamma_m\psi_A\}dy$.
\end{enumerate}
\end{theorem}

As the interior integrands $a_1^{\eta,M}$ and $a_3^{\eta,M}$ were determined previously by Branson and Gilkey
\cite{BrGi-92}, we shall concentrate upon determining the boundary integrands.

Here is a brief outline to the paper. In Section \ref{sect-2}, we
derive some basic functorial properties of these invariants. One
of the peculiarities of using the `functorial approach' is that it
is necessary to work in a very general context and then specialize
subsequently. To employ this method, we will have to work with
operators which are not self-adjoint despite the fact that the
examples which arise in practice are usually self-adjoint. In
Section \ref{sect-3}, we express the invariants
$a_n^{\eta,\partial M}$ in terms of a Weyl basis with certain
undetermined coefficients and begin the evaluation of these
coefficients. We complete the proof of Theorem \ref{thm-1.2} in
Sections \ref{sect-4} and \ref{sect-5} by completing the
determination of the coefficients.

\section{Functorial properties}\label{sect-2}
We refer to
\cite{Gi-03,GiKi-03} for the proof of the following result which
describes the adjoint structures:

\begin{lemma}\label{lem-2.1}
Let $P^*$ be the formal adjoint of $P$, let $A^*$ be the formal adjoint of $A$, and let
$A^{\#}:=\gamma_mA^*\gamma_m$.\begin{enumerate}
\item The operator $A^{\#}$ defines the adjoint boundary condition for $P^*$.
\item We have $\psi_{P^*}=\psi_P^*$, $\psi_{A^*}=\psi_A^*$, and $\psi_{A^\#}=\gamma_m\psi_A^*\gamma_m+L_{aa}\id$.
\item If $\psi_P$ is self-adjoint, and if $\psi_A=\gamma_m\psi_A^*\gamma_m+L_{aa}\id$,
then
$P_A$ is self-adjoint on $L^2(V)$.
\end{enumerate}
\end{lemma}

The next observation follows from work of Grubb and Seeley \cite{GrSe-96}.
\begin{lemma}\label{lem-2.2}
Let $n<m$. Assume that the metric on $M$ is product near the
boundary, that $P_A$ is self-adjoint, that $A$ is self-adjoint,
and that the coefficients of $P$ and of $A$ are independent of the
normal variable near the boundary. Let $F$ be an endomorphism of
$V$ whose coefficients are independent of the normal variable near
the boundary.
\begin{enumerate}
\item If $n$ is even, then
$a_n^{\zeta,\partial M}(F,P,A)= - \frac{1}{2(m-n)\Gamma( \frac{1}{2})}a_{n-1}^\eta(F,A)$.
\item If $n$ is odd, then
$a_n^{\zeta,\partial M}(F,P,A)=
    \frac{1}{4}(\beta(m-n+1)-1)
    a_{n-1}^\zeta(F,A)$.
\end{enumerate}
\end{lemma}

Taking the adjoint yields yet another useful property.

\begin{lemma}\label{lem-2.3}
Let $n<m$. Let $(P,A)$ be real operators on a real bundle $V$.
Suppose $V$ is equipped with a fiber metric. Let $P^*$ be the
formal adjoint of $P$ and let $F^*$ be the adjoint of $F$. Set
$A^{\#}=\gamma_mA^*\gamma_m$. Then $a_n^{\eta,\partial
M}(F,P,A)=a_n^{\eta,\partial M}(F^*,P^*,A^{\#})$.
\end{lemma}

\begin{proof}
As we are in the real setting, taking the complex conjugate plays no role. Consequently
\begin{eqnarray*}
\Tr_{L^2}\{FPe^{-tP_A^2}\}=\Tr_{L^2}\{F^*P^*e^{-t(P_A^*)^2}\}\,.
\end{eqnarray*}
The Lemma follows by equating powers of $t$ in the asymptotic expansions and by using Lemma \ref{lem-2.1} to see
that
$A^{\#}$ defines the adjoint boundary condition.
\end{proof}

There is a useful relation between the $\zeta$ and the $\eta$ invariants.

\begin{lemma}\label{lem-2.4}
Let $F\in C^\infty(\operatorname{End}(V))$. Let $(A,P)$ be as
above and let $n<m$.
\begin{enumerate}
\item Let
$P_\varepsilon:=P+\varepsilon F$. Then
\begin{enumerate}
\item $\partial_\varepsilon
a_n^\eta(1,P_\varepsilon,A)=(n-m)a_{n-1}^\zeta(F,P_\varepsilon,A)$.
\item
$\partial_\varepsilon a_n^\zeta(1,P_\varepsilon,A)=-2a_{n-1}^\eta(F,P_\varepsilon,A)$.
\end{enumerate}
\item Let $P_\varepsilon:=P+\varepsilon\id$. Then
\begin{enumerate}
\item $\partial_\varepsilon
a_n^\eta(F,P_\varepsilon,A)=(n-m)a_{n-1}^\zeta(F,P_\varepsilon,A)$.
\item
$\partial_\varepsilon a_n^\zeta(F,P_\varepsilon,A)=-2a_{n-1}^\eta(F,P_\varepsilon,A)$.
\end{enumerate}
\item Let $P_\varepsilon:=e^{-\varepsilon f}P$ where $f$ is a smooth scalar function vanishing on $\partial M$. Then
$\partial_\varepsilon  a_n^\eta(1,P_\varepsilon,A)=(m-n)a_n^\eta(f,P_\varepsilon,A)$.
\end{enumerate}
\end{lemma}

\begin{proof} To prove Assertion (1), let
$P_\varepsilon:=P+\varepsilon F$. We compute
\begin{eqnarray*}
&&\textstyle\sum_n\partial_\varepsilon a_n^\eta(1,P_\varepsilon,A)t^{(n-m-1)/2}
 \sim\partial_\varepsilon\Tr\{P_\varepsilon e^{-tP_{\varepsilon,A}^2}\}\\
&=&\Tr\{F(\id-2tP_\varepsilon^2)e^{-tP_{\varepsilon,A}^2}\}
  =(1+2t\partial_t)\Tr\{Fe^{-tP_{\varepsilon,A}^2}\}\\
&\sim&(1+2t\partial_t)\textstyle\sum_ka_k^\zeta(F,P_\varepsilon,A)t^{(k-m)/2}\\
&=&\textstyle\sum_k
(1+k-m)a_k^\zeta(F,P_\varepsilon,A)t^{(k-m)/2}\,.
\end{eqnarray*}
Setting $k=n-1$ and equating terms in the asymptotic expansions establishes Assertion (1a). Similarly, we compute:
\begin{eqnarray*}
&&\textstyle\sum_n\partial_\varepsilon a_n^\zeta(1,P_\varepsilon,A)t^{(n-m)/2}
\sim\partial_\varepsilon\Tr\{e^{-tP_{\varepsilon,A}^2}\}\\
&=&-2t\Tr\{FP_\varepsilon e^{-tP_{\varepsilon,A}^2}\}
\sim\textstyle\sum_k-2a_k^\eta(F,P_\varepsilon,A)t^{(k-m+1)/2}\,.
\end{eqnarray*}
Again, equating coefficients in the associated asymptotic expansions yields Assertion (1b); the proof of Assertion
(2) is similar and is therefore omitted. To prove Assertion (3), we compute:
\begin{eqnarray*}
&&\textstyle\sum_n\partial_\varepsilon
a_n^\eta(1,P_\varepsilon,A)t^{(n-m-1)/2}
\sim\partial_\varepsilon\Tr\{P_\varepsilon e^{-tP_{\varepsilon,A}^2}\}\\
&=&-\Tr\{f(P_\varepsilon-2tP_\varepsilon^3)e^{-tP_{\varepsilon,A}^2}\}
=-(1+2t\partial_t)\Tr\{fP_\varepsilon e^{-tP_{\varepsilon,A}^2}\}\\
&=&-\textstyle\sum_n(1+(n-m-1))a_n^\eta(f,P_\varepsilon,A)t^{(n-m-1)/2}\,.
\end{eqnarray*}
Assertion (3) now follows by equating coefficients in the asymptotic expansions.\end{proof}

We will need the following Lemma to apply Lemma \ref{lem-2.4}. It involves a formula for endomorphism valued
smearing functions which is related to the product case and which generalizes the formula of Theorem \ref{thm-1.1}
(3).
\begin{lemma}\label{lem-2.5}
Assume that the
metric on $M$ is product near the boundary, that $P_A$ is self-adjoint, that $A$ is self-adjoint, and that the
coefficients of
$P$ and of
$A$ are independent of the normal variable near the boundary. Let $F$ be an endomorphism of $V$ whose coefficients
are independent of the normal variable near the boundary. If $m\ge3$, then
\begin{eqnarray*}
a_2^{\zeta}(F,P,A)&=&(4\pi)^{-m/2}\textstyle\int_M\Tr\{F(\ffrac16\tau+E)\}dx\\
\hphantom{a_2^{\zeta}(F,P,A)}
&-&\ffrac1{2(m-2)}(4\pi)^{-m/2}\textstyle\int_{\partial M}
   \Tr\{(3-m)F\psi_A+F\gamma_a^T\psi_A\gamma_a^T\}dy\,.
\end{eqnarray*}
\end{lemma}

\begin{remark}\rm To ensure that $P_A$ is self-adjoint, we impose the relations of Lemma \ref{lem-2.1} (3). Since
$L_{aa}=0$ by assumption, this means that $\psi_A=\gamma_m\psi_A\gamma_m$ and hence $\Tr\{\psi_A\}=0$. Thus
$a_2^{\zeta,\partial M}(\id,P,A)=0$; this is in agreement with Theorem
\ref{thm-1.1} (3). \end{remark}

\begin{proof} We refer to \cite{BrGi-92a} for the determination of the interior integrand. Let $N=\partial M$. We
apply Theorem
\ref{thm-1.1} to the operator
$A$ on the closed manifold
$N$ to see
$$
a_2^\zeta(1,A)
=-\ffrac1{12}(4\pi)^{-(m-1)/2}\textstyle\int_N\Tr\{\tau\id+(12-6(m-1))\psi_A^2+6\psi_A\gamma_a^T\psi_A\gamma_a^T\}dy\,.
$$
We set $A_\varepsilon:=A+\varepsilon F$. By Lemma \ref{lem-2.4}, with an appropriate dimension shift,
\begin{eqnarray*}
&&-2a_1^\eta(F,A)=\partial_\varepsilon|_{\varepsilon=0}a_2^\zeta(1,A_\varepsilon)\\&=&
-\ffrac16(4\pi)^{-(m-1)/2}\textstyle\int_N\Tr\{F[(18-6m)\psi_A+6\gamma_a^T\psi_A\gamma_a^T]\}dy\,.
\end{eqnarray*}
Combining this result with Lemma \ref{lem-2.2} (1) then yields:
\medbreak\qquad $a_2^{\zeta,\partial M}(F,P,A)=-\ffrac1{2(m-2)\sqrt\pi}a_1^\eta(F,A)$
\par\quad $=-\ffrac1{12(m-2)}(4\pi)^{-m/2}\textstyle\int_{\partial
M}\Tr\{(18-6m)F\psi_A+6F\gamma_a^T\psi_A\gamma_a^T\}dy$. \end{proof}

\section{A formula with universal coefficients}\label{sect-3}

As $a_n^\eta(F,-P,A)=-a_n^\eta(F,P,A)$,
the boundary contributions, which are homogeneous of weight $n-1$, must be odd functions of $P$. Consequently,
they vanish for $n=0,1$; Assertions (1) and (2) of Theorem \ref{thm-1.2} now follow. Furthermore, we have:
\begin{lemma}\label{lem-3.1}
There exist universal constants $c_i(m)$ so that
\begin{enumerate}
\item $a_2^{\eta,\partial
M}(f,P,A)=(4\pi)^{-(m-1)/2}\textstyle\int_{\partial M}f\Tr
   \{c_m^1\psi_P+c_m^2\gamma_m\psi_A\}dy$.
\item $a_3^{\eta,\partial M}(f,P,A)=(4\pi)^{-m/2}\textstyle\int_{\partial M}\Tr\{
c_m^3f\gamma_m\psi_P^2+c_m^4f\gamma_m\gamma_a^T\psi_P\gamma_a^T\psi_P$
\smallbreak
$+c_m^5f\gamma_m\psi_A^2+c_m^6f\psi_P\psi_A+c_m^7f\gamma_m\psi_P\gamma_m\psi_A
+c_m^8f\gamma_a^T\psi_P\gamma_a^T\psi_A$
\smallbreak $+c_m^9f\gamma_a\psi_P\gamma_a\psi_A
+c_m^{10}f\gamma_m\gamma_a^T\psi_A\gamma_a^T\psi_A
+c_m^{11}f\psi_{P;m}$
\smallbreak $+c_m^{12}fL_{aa}\psi_P
+c_m^{13}f_{;m}\psi_P+c_m^{14}f(\gamma_a^T\psi_{P})_{:a}+c_m^{15}fL_{aa}\gamma_m\psi_A$
\smallbreak $+c_m^{16}f_{;m}\gamma_m\psi_A+c_m^{17}f(\gamma_a\psi_{A})_{:a}\}dy$.\end{enumerate}\end{lemma}

Many invariants do not occur because the trace over an odd number of $\gamma$-matrices is zero. Furthermore,
invariants of the form $W_{ab}\gamma_m\gamma_a\gamma_b$ and $W_{am}\gamma_a$  are omitted as their trace vanishes
as well.

We begin our study of these coefficients by varying the compatible
connection chosen:
\begin{lemma}\label{lem-3.2}
We have the relations:
\begin{enumerate}
\item $c_m^3=0$.
\item $c_m^6-c_m^7+(m-1)c_m^8+(m-1)c_m^9=0$.
\item $c_m^6+c_m^7+(m-3)c_m^8-(m-3)c_m^9+2(m-3)c_m^4=0$.
\item $c_m^6+c_m^7-(m-3)c_m^8+(m-3)c_m^9+2(m-3)c_m^{10}=0$.
\end{enumerate}\end{lemma}

\smallbreak\noindent{\it Proof.} There always exist Hermitian
connections so $\nabla\gamma=0$, see for example \cite{BrGi-92}.
There are, however, many such connections. If
$\omega:=\varrho_ie^i$ is a purely imaginary $1$ form, then
$\tilde\nabla:=\nabla-\omega\id$ is again a Hermitian connection
with $\tilde\nabla\gamma=0$. One has
$$
\tilde\psi_P=\psi_P+\varrho_i\gamma_i\quad\text{and}\quad
  \tilde\psi_A=\psi_A+\varrho_b\gamma_b^T\,.
$$
Clearly $a_n^\eta$ does not depend on the particular connection chosen. We exhibit the terms which are linear in
$\varrho$ and omit the remaining terms to derive the following equations from which the desired relations of the
Lemma will follow:
\begin{eqnarray*}
&&\Tr\{c_m^3\gamma_m\tilde\psi_P^2\}=-2c_m^3\varrho_m\Tr\{\psi_P\}+...,\\
&&\Tr\{c_m^4\gamma_m\gamma_a^T\tilde\psi_P\gamma_a^T\tilde\psi_P\}
   =c_m^4\Tr\{-2(m-3)\gamma_m\varrho_b\gamma_b\psi_P\}+...,\\
&&\Tr\{c_m^5\gamma_m\tilde\psi_A^2\}=0+...,\\
&&\Tr\{c_m^6\tilde\psi_P\tilde\psi_A\}=c_m^6\Tr\{\varrho_m\gamma_m\psi_A
    +\varrho_b\gamma_b\psi_A+\psi_P\varrho_b\gamma_b^T\}+...,\\
&&\Tr\{c_m^7\gamma_m\tilde\psi_P\gamma_m\tilde\psi_A\}
   =c_m^7\Tr\{-\varrho_m\gamma_m\psi_A+\varrho_b\gamma_b\psi_A+
\psi_P\varrho_b\gamma_b^T\}+...,\\
&&\Tr\{c_m^8\gamma_a^T\tilde\psi_P\gamma_a^T\tilde\psi_A\}
   =c_m^8\Tr\{(m-1)\varrho_m\gamma_m\psi_A-(m-3)\varrho_b\gamma_b\psi_A\\
&&\qquad\qquad\qquad\qquad\qquad
  +(m-3)\psi_P\varrho_b\gamma_b^T\}+...,\\
&&\Tr\{c_m^9\gamma_a\tilde\psi_P\gamma_a\tilde\psi_A\}=c_m^9\Tr\{(m-1)\varrho_m\gamma_m\psi_A
+(m-3)\varrho_b\gamma_b\psi_A\\
&&\qquad\qquad\qquad\qquad\qquad
   -(m-3)\psi_P\varrho_b\gamma_b^T\}+...,\\
&&\Tr\{c_m^{10}\gamma_m\gamma_a^T\tilde\psi_A\gamma_a^T\tilde\psi_A\}
   =c_m^{10}\Tr\{2(m-3)\gamma_m\varrho_b\gamma_b^T\psi_A\}+...\qquad\quad\qedbox\\
\end{eqnarray*}

We shift the spectrum of $A$ to show:
\begin{lemma}\label{lem-3.3}
We have the relations:
\begin{enumerate}
\item $c_m^5=0$.
\item $c_m^6=c_m^7$ and $c_m^8=-c_m^9$.
\end{enumerate}
\end{lemma}

\begin{proof} If we replace $A$ by $A+\varepsilon\id$, then the boundary
condition is unchanged for small values of $\varepsilon$. We set $\tilde\psi_A:=\psi_A+\varepsilon\id$, exhibit only
the linear terms, and omit all terms which are not linear in $\varepsilon$ to derive the following equations:
\begin{eqnarray*}
&&\Tr\{c_m^5\gamma_m\tilde\psi_A^2\}=2c_m^5\Tr\{\gamma_m\varepsilon\psi_A\}+...,\\
&&\Tr\{c_m^6\psi_P\tilde\psi_A\}=c_m^6\Tr\{\varepsilon\psi_P\}+...,\\
&&\Tr\{c_m^7\gamma_m\psi_P\gamma_m\tilde\psi_A\}=-c_m^7\Tr\{\varepsilon\psi_P\}+...,\\
&&\Tr\{c_m^8\gamma_a^T\psi_P\gamma_a^T\tilde\psi_A\}=c_m^8\Tr\{-(m-1)\varepsilon\psi_P\}+...,\\
&&\Tr\{c_m^9\gamma_a\psi_P\gamma_a\tilde\psi_A\}=c_m^9\Tr\{-(m-1)\varepsilon\psi_P\}+...,\\
&&\Tr\{c_m^{10}\gamma_m\gamma_a^T\tilde\psi_A\gamma_a^T\tilde\psi_A\}=0+...
\end{eqnarray*}
Assertion (1) follows. Furthermore, we have
$$0=c_m^6-c_m^7-(m-1)c_m^8-(m-1)c_m^9\,.$$
Assertion (2) follows from this equation and from Lemma
\ref{lem-3.2} (2).\end{proof}

\begin{lemma}\label{lem-3.4}
We have $c_m^{14}=0$ and $c_m^{17}=0$.
\end{lemma}

\begin{proof} We work on the flat annulus $M:=\mathbb{T}^{m-1}\times[0,1]$. Let $h_a$ and $H_a$ be real
smooth functions
on $M$. We set
$$P=\gamma_i\partial_i^x+\varepsilon h_a\gamma_a^T\quad\text{and}\quad
A=\gamma_a^T\partial_a^y+\varepsilon H_b\gamma_b\,.
$$
Let $F=f\cdot\id$ be scalar. The presence of the smearing function
$f$ ensures the boundary and interior integrals do not interact.
Modulo terms which are $O(\varepsilon^2)$, one has:
$$a_3^\eta(F,P,A)=-\varepsilon(4\pi)^{-m/2}\textstyle\int_{\partial M}
\mbox{Tr}\left(
f\{c_m^{14}h_{b:b}+c_m^{17}H_{b:b}\}\right)dy+O(\varepsilon^2)\,.$$
By Lemma \ref{lem-2.1},
$$\begin{array}{ll}
P^*=\gamma_i\partial_i^x-\varepsilon h_a\gamma_a^T,&\psi_{P^*}=-\varepsilon h_a\gamma_a^T,\\
A^{\#}=\gamma_mA^*\gamma_m=\gamma_m(\gamma_a^T\partial_a^y-\varepsilon
H_b\gamma_b)\gamma_m,&\psi_{A^{\#}}=-\varepsilon H_b\gamma_b\,.
\end{array}$$
Consequently, there is a sign change
$$a_3^\eta(F,P^*,A^{\#})=\varepsilon(4\pi)^{-m/2}\textstyle\int_{\partial M}
\mbox{Tr}\left(
f\{c_m^{14}h_{b:b}+c_m^{17}H_{b:b}\}\right)dy+O(\varepsilon^2)\,.$$
By Lemma  \ref{lem-2.3},
$a_3^\eta(f\id,P^*,A^{\#})=a_3^\eta(f\id,P,A)$; the Lemma follows.
\end{proof}

We use conformal variations to show:
\begin{lemma}\label{lem-3.5}
$c_m^{15}=\frac{m-3}{1-m}c_m^{16}$.
\end{lemma}

\begin{proof} Let $f$ be a smooth function with $f|_{\partial M}=0$. Let
$ds^2(\varepsilon)=e^{2\varepsilon f}ds^2$ and let $P(\varepsilon):=e^{-\varepsilon f}P$.
Let $\nabla$ be a unitary connection with $\nabla\gamma=0$. Let $x=(x_1,...,x_m)$ be a system of local
coordinates on $M$. Expand $P=\gamma^\nu\nabla_{\partial_\nu}+\psi_P$ and use the metric to lower indices and
define $\gamma_\nu$. Define a smooth $1$ parameter family of connections
$$\nabla(\varepsilon)_{\partial_\mu}:=\nabla_{\partial_\mu}+\ffrac\varepsilon2\{f_{;\nu}\gamma^{\nu}\gamma_{\mu}
   +f_{;\mu}\}\,.$$
Results of \cite{DoGiKi99} show $\nabla(\varepsilon)\gamma(\varepsilon)=0$ and $\nabla(\varepsilon)$ is unitary.
Furthermore,
$$\psi_P(\varepsilon)=e^{-\varepsilon f}\{\psi_P-\ffrac{m-1}2\varepsilon f_{;i}\gamma_i\}
\quad\text{and}\quad\psi_A(\varepsilon)=\psi_A\,.$$
We suppose $\psi_P=0$. We study the term $\Tr\{f_{;m}\gamma_m\psi_A\}$ and compute:
\begin{eqnarray*}
&&\partial_\varepsilon|_{\varepsilon=0}\Tr\{c_m^6\psi_P\psi_A+c_m^7\gamma_m\psi_P\gamma_m\psi_A\}
   \\&&\qquad=-\ffrac{m-1}2(c_m^6-c_m^7)\Tr\{f_{;m}\gamma_m\psi_A\}=0,\\
&&\partial_\varepsilon|_{\varepsilon=0}\Tr\{c_m^8\gamma_a^T\psi_P\gamma_a^T\psi_A
    +c_m^9\gamma_a\psi_P\gamma_a\psi_A\}\\
  &&\qquad=-\ffrac{(m-1)^2}2\{c_m^8+c_m^9\}\Tr\{f_{;m}\gamma_m\psi_A\}=0,\\
&&\partial_\varepsilon|_{\varepsilon=0}L_{aa}=(1-m)f_{;m}\,.
\end{eqnarray*}
We concentrate on the term $\Tr\{f_{;m}\gamma_m\psi_A\}$ and compute
\begin{eqnarray*}
&&\partial_\varepsilon|_{\varepsilon=0}a_3^\eta(1,P(\varepsilon),A)=
(4\pi)^{-m/2}\textstyle\int_{\partial M}c_m^{15}\Tr\{(1-m)f_{;m}\gamma_m\psi_A\}dy\\
&=&(m-3)a_3^\eta(f,P(\varepsilon),A)
  =(4\pi)^{-m/2}\textstyle\int_{\partial M}c_m^{16}\Tr\{(m-3)f_{;m}\gamma_m\psi_A\}dy\,.
\end{eqnarray*}
The Lemma now follows.
\end{proof}

We study a variation of the form $P_\varepsilon:=P+\varepsilon\id$ to establish
\begin{lemma}\label{lem-3.6}\
\begin{enumerate}
\item $c_m^1=\ffrac{2-m}4(\beta(m)-1)$.
\item $c_m^{12}=\ffrac{3-m}3(1-\ffrac34\pi\beta(m))$ and
      $c_m^{13}=\ffrac{(m-3)(m-1)}{2(m-2)}(1-\ffrac12\pi\beta(m))$.
\end{enumerate}\end{lemma}

\begin{proof} Let $\psi_P$ be self-adjoint. Set
$\psi_A:=\frac12L_{aa}\id$; then $A^{\#}=A^*=A$ and $P_A$ is self-adjoint. Let $P_\varepsilon:=P+\varepsilon\id$.
By Theorem \ref{thm-1.1} and Lemma
\ref{lem-2.4}:
\begin{eqnarray*}
&&\partial_\varepsilon|_{\varepsilon=0}a_2^{\eta,\partial M}(f,P_\varepsilon,A)
=(4\pi)^{-(m-1)/2}\textstyle\int_Mc_m^1f\Tr\{\id\}dy\\
&=&\textstyle(2-m)a_1^{\zeta,\partial M}(f,P,A)=(4\pi)^{-(m-1)/2}\ffrac{2-m}4(\beta(m)-1)\int_{\partial
M}f\Tr\{\id\}dy
\end{eqnarray*}
Assertion (1) follows. To establish Assertion (2), we compute:
\begin{eqnarray*}
&&\Tr\{c_m^6\psi_P\psi_A+c_m^7\gamma_m\psi_P\gamma_m\psi_A\}\\
&&\qquad=\ffrac12(c_m^6-c_m^7)\Tr\{\psi_PL_{aa}\}=0,\\
&&\Tr\{c_m^8\gamma_a^T\psi_P\gamma_a^T\psi_A+c_m^9\gamma_a\psi_P\gamma_a\psi_A\}\\
&&\qquad=\ffrac{(1-m)}2(c_m^8+c_m^9)\Tr\{\psi_PL_{aa}\}=0,\\
&&\partial_\varepsilon|_{\varepsilon=0}c_m^4\Tr\{\gamma_m\gamma_a^T\psi_P\gamma_a^T\psi_P)\\
&&\qquad=c_m^4\Tr(\gamma_m\gamma_a^T\gamma_a^T\psi_P+\gamma_a^T\gamma_m\gamma_a^T\psi_P)=0\,.
\end{eqnarray*}
Consequently again by Theorem \ref{thm-1.1} and Lemma
\ref{lem-2.4} one has:
\begin{eqnarray*}
&&\partial_\varepsilon|_{\varepsilon=0}a_3^{\eta,\partial M}(f,P_\varepsilon,A)
=(4\pi)^{-m/2}\textstyle\int_{\partial M}\Tr\{c_m^{13}f_{;m}\id+c_m^{12}fL_{aa}\}dy\\
&=&(3-m)a_2^{\zeta,\partial M}(f,P,A)\\
&=&(3-m)\textstyle\int_{\partial M}\{\frac13(1-
     \frac34\pi \beta(m))L_{aa} f
     -\frac{m-1}{2(m-2)}(1-
        \frac12 \pi \beta(m))f_{;m}\}\Tr\{\id\}dy\,.
\end{eqnarray*}
Assertion (2) follows.\end{proof}

\section{The variation $P_\varepsilon:=P+\varepsilon F$}\label{sect-4}

In this section, we will study $\partial_\varepsilon
a_3^\eta(1,P_\varepsilon,A)$. There is a non-trivial interaction
between the boundary and interior integrals that must be dealt
with. Our basic identity is provided by Lemma \ref{lem-2.4},
\begin{equation}\label{eqn-4.a}
\partial_\varepsilon|_{\varepsilon=0}a_3^\eta(1,P_\varepsilon,A)=(3-m)a_2^\zeta(F,P,A)\,.
\end{equation}
Let $F$ be endomorphism valued. Then:
\begin{eqnarray*}
&&\partial_\varepsilon|_{\varepsilon=0}a_3^{\eta,M}(1,P_\varepsilon,A)=-\ffrac1{12}(4\pi)^{-m/2}
\textstyle\int_M\Tr\{[2(m-1)F_{;i}+3(4-m)F\gamma_i\psi_P\\
&&\quad+3(4-m)F\psi_P\gamma_i+3F\gamma_j\gamma_i\psi_P\gamma_j
+3F\gamma_j\psi_P\gamma_j\gamma_i]_{;i}\\
&&\quad+(3-m)[F\tau+6F\gamma_i\gamma_jW_{ij}-6F\psi_{P;i}\gamma_i-6F_{;i}\gamma_i\psi_P
+3(4-m)F\psi_P\psi_P\\
&&\quad+3F\psi_P\gamma_i\psi_P\gamma_i
+3F\gamma_i\psi_P\gamma_i\psi_P+3F\gamma_i\psi_P\psi_P\gamma_i]\}dx\,.
\end{eqnarray*}
On the other hand, by Lemma \ref{lem-2.5},
\begin{eqnarray*}
&&a_2^{\zeta,M}(F,P,A)=-\ffrac1{12}(4\pi)^{-m/2}\textstyle\int_M\Tr\{F
(\tau+6\gamma_i\gamma_jW_{ij}+6\gamma_i\psi_{P;i}
-6\psi_{P;i}\gamma_i\\&&\qquad+12\psi_P^2+3\psi_P\gamma_i\psi_P\gamma_i
+3\gamma_i\psi_P\psi_P\gamma_i+3\gamma_i\psi_P\gamma_i\psi_P
-3m\psi_P^2)\}dx\,.
\end{eqnarray*}
Consequently, we may integrate by parts to see
\begin{eqnarray}\label{eqn-4.b}
&&\partial_\varepsilon
a_3^{\eta,M}(1,P_\varepsilon,A)|_{\varepsilon=0}-(3-m)a_2^{\zeta,M}(F,P,A)\\ &=&-\ffrac1{12}(4\pi)^{-m/2}
\textstyle\int_M\Tr\{[2(m-1)F_{;i}+3(4-m)F\gamma_i\psi_P+3(4-m)F\psi_P\gamma_i\nonumber\\
&&+3F\gamma_j\gamma_i\psi_P\gamma_j
+3\gamma_j\psi_P\gamma_j\gamma_iF]_{;i}
-6(3-m)F_{;i}\gamma_i\psi_P-6(3-m)F\gamma_i\psi_{P;i}\}dx\nonumber\\
&=&\ffrac1{12}(4\pi)^{-m/2}
\textstyle\int_{\partial M}\Tr\{2(m-1)F_{;m}+3(4-m)F\gamma_m\psi_P+3(4-m)F\psi_P\gamma_m\nonumber\\
&&+3F\gamma_j\gamma_m\psi_P\gamma_j
+3F\gamma_j\psi_P\gamma_j\gamma_m-6(3-m)F\gamma_m\psi_P\}dy\,.\nonumber
\end{eqnarray}
After setting $c_m^3=c_m^5=0$, $c_m^7=c_m^6$, and $c_m^9=-c_m^8$, one has
\begin{eqnarray}\label{eqn-4.c}
&&\partial_\varepsilon a_3^{\eta,\partial
M}(1,P_\varepsilon,A)|_{\varepsilon=0}=
(4\pi)^{-m/2}\textstyle\int_{\partial M}\Tr\{
c_m^4F(\gamma_a^T\psi_P\gamma_m\gamma_a^T+\gamma_m\gamma_a^T\psi_P\gamma_a^T)\\
&&\quad+c_m^6F(\psi_A+\gamma_m\psi_A\gamma_m)+c_m^8F(\gamma_a^T\psi_A\gamma_a^T-\gamma_a\psi_A\gamma_a)
+c_m^{11}F_{;m}\nonumber\\
&&\quad+c_m^{12}FL_{aa}
\}dy\,.\nonumber
\end{eqnarray}

There are several different settings where we know
$a_2^{\zeta,\partial M}$. For the next two lemmas, to ensure that $P_A$ is self adjoint, we shall assume
$\psi_P$ and $\psi_A$ are self adjoint and that
$\psi_A=\gamma_m\psi_A\gamma_m+L_{aa}\id$. We begin by applying
Theorem \ref{thm-1.1}:

\begin{lemma}\label{lem-4.1}
We have $c_m^{11}=\ffrac{(m-3)(m-1)}{2(m-2)}(1-\ffrac12\pi\beta(m))-\frac16(m-1)$.
\end{lemma}

\begin{proof} We take $F=f\cdot\id$ to be scalar and set $P_\varepsilon:=P+\varepsilon F$. The terms involving
$\Tr(\psi_A)$ and $\Tr(\gamma_m\psi_P)$ cancel and we have
\begin{eqnarray*}
0&=&\partial_{\varepsilon}a_3^\eta(1,P_\varepsilon,A)|_{\varepsilon=0}-(3-m) a_2^\zeta(F,P,A)\\
&=&\Tr\{\id\}(4\pi)^{-m/2}\textstyle\int_{\partial M}\{\ffrac16(m-1)+c_m^{11})f_{;m}
+c_m^{12}fL_{aa}\\
&-&\textstyle\ffrac{3-m}3(1-
     \frac34\pi \beta(m))L_{aa} f
     +\frac{(m-1)(3-m)}{2(m-2)}(1-
        \frac12 \pi \beta(m))f_{;m}\}dy\,.
\end{eqnarray*}
We equate the coefficients of $fL_{aa}$ to determine a value for $c_m^{12}$ which agrees with that obtained
in Lemma \ref{lem-3.6}. Equating the coefficients of $f_{;m}$ determines $c_m^{11}$. \end{proof}


We apply Lemma \ref{lem-2.5} to prove:
\begin{lemma}\label{lem-4.2}
We have the relations:
\begin{enumerate}
\item $c_m^6=-\ffrac{(3-m)^2}{4(m-2)}$, and $c_m^8=-\ffrac{3-m}{4(m-2)}$.
\item $c_m^4=0$, and $c_m^{10}=-2\ffrac{(3-m)}{4(m-2)} .$
\end{enumerate}
\end{lemma}

\begin{proof} We assume the structures are product near the boundary. We first study the terms
$\Tr\{F\psi_A\}$ and $\Tr\{\gamma_aF\gamma_a\psi_A\}$. Since
$L_{aa}=0$, $\gamma_m\psi_A\gamma_m=\psi_A$. We compute using
equations (\ref{eqn-4.b}) and (\ref{eqn-4.c}) that
\begin{eqnarray*}
&&c_m^6\partial_\varepsilon|_{\varepsilon=0}
\Tr\{\psi_P\psi_A+\gamma_m\psi_P\gamma_m\psi_A\}=2c_m^6\Tr\{F\psi_A\},\\
&&c_m^8\partial_\varepsilon|_{\varepsilon=0}\Tr\{\gamma_a^T\psi_P\gamma_a^T\psi_A-\gamma_a\psi_P\gamma_a\psi_A\}
    =-2c_m^8\Tr\{\gamma_aF\gamma_a\psi_A\}\,.
\end{eqnarray*}
Thus Lemma \ref{lem-2.5} yields:
\begin{eqnarray}\label{eqn-4.d}
&&(4\pi)^{-m/2}\textstyle\int_{\partial M}\Tr\{2c_m^6F\psi_A
-2c_m^8F\gamma_a\psi_A\gamma_a\}dy+...\\
&=&-\ffrac{3-m}{2(m-2)}(4\pi)^{-m/2}
   \textstyle\int_{\partial M}\Tr\{F(3-m)\psi_A-F\gamma_a\psi_A\gamma_a\}dy+...\,.\nonumber
\end{eqnarray}
To complete the proof of Assertion (1), we must show Equation
(\ref{eqn-4.d}) yields two linearly independent relations. If we
set $F=\psi_A=\sqrt{-1}\gamma_1$, then $\psi_A^*=\psi_A$,
$\gamma_m\psi_A\gamma_m=\psi_A$, and
$$\Tr(F\psi_A)=\Tr\{\id\}\quad\text{and}\quad\Tr(F\gamma_a\psi_A\gamma_a)=(m-3)\Tr\{\id\}\,.$$
If we set $F=\psi_A=\gamma_1\gamma_2\gamma_3$, then $\psi_A^*=\psi_A$, $\gamma_m\psi_A\gamma_m=A$, and
$$
\Tr(F\psi_A)=\Tr\{\id\}\quad\text{and}\quad\Tr(F\gamma_a\psi_a\gamma_a)=(m-7)\Tr\{\id\}\,.
$$
Assertion (1) follows. Assertion (1) and Lemma \ref{lem-3.2} imply Assertion (2).
\end{proof}

\section{A special case calculation on the ball}\label{sect-5}
In order to find the remaining unknown coefficients $c_m^2$ and
$c_m^{16}$ we evaluate the leading coefficients in the asymptotic
of the eta invariant for an example on the ball. We first describe
the setting considered.

Let $r\in[0,1]$ be the radial normal coordinate and $d\Sigma^2$
the usual metric on the unit sphere $S^{m-1}$. Then the standard
metric on the ball is $ds^2=dr^2+r^2d\Sigma^2$. The inward unit
normal on the boundary is $-\partial_r$. For this metric, the only
nonvanishing components of the Christoffel symbols are
\begin{eqnarray}
\Gamma_{abc}= \frac 1 r \tilde{\Gamma}_{abc}\text{ and }
\Gamma_{abm}= \frac 1 r \delta_{ab};\nn
\end{eqnarray}
the second fundamental form is given by $L_{ab}=\delta_{ab}$. We
will use $\tilde{\Gamma}_{abc}$ to refer to the Christoffel
symbols associated with the metric $d\Sigma^2$ on the sphere
$S^{m-1}$. We will consider the Dirac operator
$P=\gamma^\nu\partial_\nu$ on the ball; we take the flat
connection $\nabla$ and set $\psi_P=0$. We suppose $m$ even (there
is a corresponding decomposition for $m$ odd) and use the
following representation of the $\gamma$-matrices,
\begin{eqnarray}
&&\gamma_{a(m)}=\left(
   \begin{array}{cc}
               0 &  \sqrt{-1}\cdot \gamma_{a(m-1)}    \\
      -\sqrt{-1}\cdot \gamma_{a(m-1)}    &     0
    \end{array}    \right)\text{ and }\nn\\
\quad &&\gamma_{m(m)} = \left(
     \begin{array}{cc}
         0       &    \sqrt{-1}\cdot 1_{m-1}   \\
    \sqrt{-1}\cdot 1_{m-1}\quad\    &      0
    \end{array}   \right)   .\nonumber
\end{eqnarray}
We stress that the matrices $\gamma_{j(m)}$ are the
$\gamma$-matrices projected along some vielbein system $e_j$. We
decompose $\nabla_j = e_j + \omega_j$ where $\omega_j=\frac 1 4
\Gamma_{jkl} \gamma_{k(m)} \gamma_{l(m)}$ is the connection-$1$
form of the spin connection. If $\tilde{\nabla}$ denotes the
connection on the sphere, we have that
$$
\nabla_a = \frac 1 r\left( \left(
        \bea {cc}
        \tilde{\nabla}_a & 0 \\
         0 & \tilde{\nabla}_a
         \eea  \right)  +\frac 1 2 \gamma_{a(m)}^T\right).
$$
This allows us to decompose the Dirac operator on the ball into a
radial part and a part living on the sphere. In detail, if $\tilde
P$ is the Dirac operator on the sphere, we have
\begin{eqnarray}
&&P=\left(\frac{\partial}{\partial x_m}-\frac{m-1}{2r} \right)
\gamma_{m(m)}
         +\frac 1 r \left(
\begin{array}{cc}
       0   & \sqrt{-1} \tilde P \\
    -\sqrt{-1} \tilde P  & 0
\end{array}   \right).\nn
\end{eqnarray}
Let $d_s$ be the dimension of the spin bundle on the disk;
$d_s=2^{m/2}$ if $m$ is even. The spinor modes ${\mathcal Z} _\pm
^{(n)}$ on the sphere are discussed in \cite{roberto}. We have
\begin{eqnarray*}
&&\tilde P {\mathcal Z} _\pm ^{(n)} (\Omega )=  \pm \left(
n+\frac{m-1} 2 \right)
             {\mathcal Z} _\pm ^{(n)} (\Omega )\text{ for }n=0,1,...;\\
&&  d_n(m):=\dim {\mathcal Z} _\pm ^{(n)} (\Omega )= \frac 1 2 d_s
\left(
    \bea {c}
       m+n-2 \\
        n
      \eea \right) .
\end{eqnarray*}
Let $J_{\nu} (z)$ be the Bessel functions. These satisfy the
differential equation \cite{grad}:
\begin{eqnarray*}
\frac{d^2 J_{\nu} (z)} {dz^2} +\frac 1 z \frac{dJ_{\nu} (z)} {dz}
+ \left( 1-\frac{\nu^2}{z^2} \right) J_{\nu }(z) =0.
\end{eqnarray*}
Let  $P\varphi_\pm = \pm \mu \varphi_\pm$ be an eigen function of
$P$. Modulo a suitable radial normalizing constant $C$, we may
express:
\begin{eqnarray}
\varphi_{\pm}^{(+)}&=&{\frac{C}{r^{(m-2)/2}}} \left(
     \begin{array}{c}
        \sqrt{-1} J_{n+m/2}(\mu r)
       \,Z^{(n)}_+(\Omega ) \\
     \pm J_{n+m/2-1}(\mu r)\,Z^{(n)}_+(\Omega )
        \end{array}  \right)  ,\text{ and} \label{eq2.59}\\
\varphi_{\pm}^{(-)}&=&{\frac{C}{r^{(m-2)/2}}}\left(
     \begin{array}{c}
     \pm J_{n+m/2-1}(\mu r)\,Z^{(n)}_
-(\Omega )  \\
   \sqrt{-1} J_{n+m/2}(\mu r)\,Z^{(n)}_-(\Omega ) \end{array}
\right). \label{solutions}
\end{eqnarray}
We next impose the boundary conditions. We choose for
$\epsilon\in\reals$ the boundary endomorphism
 \beq \psi_A = \epsilon \gamma_{m(m)} + \frac 1
2 L_{aa}\id\label{psiA}\eeq such that \beq\psi_A = \gamma_{m(m)}
\psi_A^* \gamma_{m(m)} + L_{aa}\id.\nn\eeq This guarantees that $P_A$
is self-adjoint, see Lemma \ref{lem-2.1}, Assertions (2) and (3).
For this setting the general form of the leading coefficients for
the eta invariant are obtained from Lemma \ref{lem-3.1} and Lemma
\ref{lem-3.5}. Noting that the volume of the $m-1$ dimensional
sphere is $2\pi^{m/2}/\Gamma (m/2)$, and that \beq \mbox{Tr}
\left\{ \gamma_m \psi_A \right\}= - \epsilon d_s,\nn\eeq one finds
\beq a_2^\eta (1,P,A) &=&- c_m^2\frac{ \epsilon d_s \sqrt
\pi}{2^{m-2} \Gamma \left( \frac m 2 \right) } , \label{cm2}\\
a_3^\eta (1,P,A) &=&c_m^{16} \frac{ (m-3) \epsilon d_s}{2^{m-1}
\Gamma \left( \frac m 2 \right) } .\label{cm16}\eeq Thus finding
explicit answers for this example will allow us to determine
$c_m^2$ and $c_m^{16}$. We proceed towards this goal.

For the $\psi_A$ given in (\ref{psiA}) the boundary operator $A$
is given by \beq A =\left(
\begin{array}{cc}
-\tilde P & i\epsilon \\
i\epsilon & \tilde P \end{array} \right) .\nn\eeq We need to find
the spectral projection on those eigen spinors of $A$ whose
eigenvalues have a positive real part. The endomorphism $\psi _A$
chosen allows us to obtain closed forms for all eigenvalues
$\pm\mu_n$ and eigenspinors $(\alpha_1^\pm,\alpha_2^\pm)$ defined
by the differential equation \beq A {\alpha_1^\pm \choose \alpha_
2^\pm} = \left( \bea{cc} -\tilde P &
i\epsilon \\
i\epsilon & \tilde P \eea \right) {\alpha _1^\pm \choose
\alpha_2^\pm } = \pm\mu _n{\alpha _1^\pm \choose \alpha _2^\pm } .
\nn\eeq Let
$$\lambda _n = n + \frac 1 2 (m-1)$$ be the eigenvalues associated
with $\epsilon =0$ \cite{GiKi-03}. One can then show that $$\mu_n
= \sqrt{\lambda_n^2 -\epsilon^2}$$ and \beq \left(\bea{c}
\alpha_1^+
\\ \alpha _2^+ \eea \right) &=& \left( \bea{c} \frac
{\sqrt{-1}\epsilon}{2\lambda_n} Z_+^{(n)} + Z_-^{(n)} \\ \frac 1
{2\lambda_n} \left( \sqrt{ \lambda_n^2 -\epsilon^2} + \lambda_n
\right)  Z_+^{(n)} + \frac 1 {\sqrt{-1}\epsilon}\left( \sqrt{
\lambda_n^2 -\epsilon^2} - \lambda_n
\right) Z_-^{(n)}  \eea \right) , \nn\\[.3cm]
\left(\bea{c} \alpha_1^- \\ \alpha_2^-\eea \right) &=&
\left(\bea{c} Z_+^{(n)} - \frac{\sqrt{-1}\epsilon}{2\lambda_n} Z_-
^{(n)}
\\
-\frac 1 {\sqrt{-1}\epsilon} \left( \sqrt{\lambda_n^2 -\epsilon^2}
-\lambda_n \right) Z_+ ^{(n)} + \frac 1 {2\lambda_n} \left(
\sqrt{\lambda_n^2 - \epsilon^2} + \lambda_n \right) Z_- ^{(n)}
\eea\right). \nn\eeq We choose $\epsilon < (m-1)/2$ such that all
eigenvalues $\mu_n$ are real. The solutions are normalized such
that in the limit $\epsilon \to 0$ they reduce to the previously
determined solutions in \cite{GiKi-03}.

We want to suppress the projection on the positive spectrum of
$A$. Using the solutions given in Equations (\ref{eq2.59}) and
(\ref{solutions}) this is easily accomplished. Projecting
$\varphi_\pm^{(+)}$ onto the positive spectrum of $A$ gives the
implicit eigenvalue condition \beq J_{\lambda_n -\frac 1 2} (\mu )
\mp \frac \epsilon {\sqrt{\lambda_n^2 - \epsilon ^2} + \lambda_n }
J_{\lambda_n + \frac 1 2 } (\mu ) = 0, \nn\eeq whereas projecting
$\varphi_\pm ^{(-)}$ produces \beq J_{\lambda_n -\frac 1 2} (\mu )
\pm \frac 1 \epsilon \left( {\sqrt{\lambda_n^2 - \epsilon ^2} -
\lambda_n } \right) J_{\lambda_n + \frac 1 2 } (\mu ) = 0. \nn\eeq
Combining the equations for the positive eigenvalues of $P_A$, we
have the condition \beq & &\left(J_{\lambda_n -\frac 1 2} (\mu ) -
\frac \epsilon {\sqrt{\lambda_n^2 - \epsilon ^2} + \lambda_n }
J_{\lambda_n + \frac 1 2 } (\mu )
\right)\times\nn\\
& &\quad\quad\quad\quad\left(J_{\lambda_n -\frac 1 2} (\mu ) +
\frac 1 \epsilon \left( {\sqrt{\lambda_n^2 - \epsilon ^2} -
\lambda_n }\right) J_{\lambda_n + \frac 1 2 } (\mu )\right)
 =0.\nn\eeq For the present purpose it will be sufficient to find the unknown
multipliers $c_m^2$ and $c_m^{16}$ multiplying a linear term in
$\psi_A$. Therefor we only need to pick up linear terms in
$\epsilon$ and we will consider only terms up to the order
$\epsilon$ explicitly. Having that in mind we write the implicit
eigenvalue condition for positive eigenvalues instead as\beq
J_{\lambda_n -\frac 1 2} (\mu ) \left( J_{\lambda_n -\frac 1 2}
(\mu ) - \frac \epsilon {\lambda_n} J_{\lambda_n + \frac 1 2} (\mu
) \right) + {\mathcal O} (\epsilon ^2 ) =0.\label{impli1}\eeq To
simplify the notation, set
$$p=\lambda_n -\frac 1 2\quad\text{and}\quad d_n (m)= d_p(m)\,.$$ Furthermore, we use
the recursion for Bessel functions, see \cite{grad},
$$z\frac d {dz} J_p (z) - p J_p (z) = - z J_{p+1} (z), $$
to rewrite (\ref{impli1}) such that only the index $p$ appears,
\beq J_p (\mu ) \left( J_p (\mu ) \left[ 1 - \frac{ \epsilon
p}{\mu (p+1/2)} \right] + \frac \epsilon {p+1/2} J_p ' (\mu )
\right) + {\mathcal O} (\epsilon^2) = 0 .\label{impli2}\eeq
Proceeding similarly with the negative eigenvalues of $P_A$ the
outcome is \beq J_p (\mu ) \left( J_p (\mu ) \left[ 1 + \frac{
\epsilon p}{\mu (p+1/2)} \right] - \frac \epsilon {p+1/2} J_p '
(\mu ) \right) + {\mathcal O} (\epsilon^2) = 0 .\label{impli3}\eeq
Using Cauchy's residue theorem these equations allow us to rewrite
the eta function \beq \eta (s;P,A ) = \sum_\mu (\mbox{sign} (\mu
)) | \mu | ^{-s} \nn\eeq in terms of a contour integral, a
technique recently described in detail in
\cite{BSW02,bek,cmp,kir01}. The coefficients in the asymptotic
expansion (\ref{eqn-1.a}) are then determined by evaluating
residues of $\eta$ according to \cite{Gb} \beq \mbox{Res }\eta
(m-n; P,A) = \frac{ 2 a_n^\eta (1,P,A)}{\Gamma \left( \frac{
m-n+1} 2 \right)}.\label{etaeq}\eeq We will need the residues at
$s=m-2$ and $s=m-3$ in order to determine the coefficients
$a_2^\eta$ and $a_3^\eta$.

Neglecting systematically the higher order terms in $\epsilon$, we
use a suitable counterclockwise contour $C$ enclosing all the
solutions of the Equations (\ref{impli2}) and (\ref{impli3}) to
write the eta function as (from now on it will be understood that
this is the eta function up to the order $\epsilon$) \beq \eta
(s;P,A ) &=& \sum_p d_p(m) \int\limits_C \frac {dk}{2\pi i} k^{-s}
\frac
\partial
{\partial k} \nn\\
& &\left\{ \ln \left[J_p (k ) \left( J_p (k ) \left[ 1 - \frac{
\epsilon p}{k (p+1/2)} \right] + \frac \epsilon {p+1/2}
J_p ' (k ) \right)\right] \right.\nn\\
& &\left. -\ln\left[J_p (k ) \left( J_p (k ) \left[ 1 + \frac{
\epsilon p}{k (p+1/2)} \right] - \frac \epsilon {p+1/2} J_p '
(k ) \right)\right]\right\}\nn\\
&=&\sum_p d_p(m) \int\limits_C \frac {dk}{2\pi i} k^{-s} \frac
\partial
{\partial k} \nn\\
& &\left\{ \ln \left[1- \frac{ \epsilon p}{ k(p+1/2)}+ \frac
\epsilon {p+1/2}
\frac{J_p ' (k )}{J_p (k)} \right] \right.\nn\\
& &\left. -\ln\left[ 1 + \frac{ \epsilon p}{k(p+1/2)} - \frac
\epsilon {p+1/2} \frac{J_p ' (k )}{J_p (k)}
\right]\right\}\nn\\
&=&\sum_p d_p(m) p^{-s}\int\limits_C \frac {dz}{2\pi i} z^{-s}
\frac
\partial
{\partial z} \nn\\
& &\left\{ \ln \left[1- \frac{ \epsilon p}{ zp(p+1/2)}+ \frac
\epsilon {p+1/2}
\frac{J_p ' (zp )}{J_p (zp)} \right] \right.\nn\\
& &\left. -\ln\left[ 1 + \frac{ \epsilon p}{zp(p+1/2)} - \frac
\epsilon {p+1/2} \frac{J_p ' (zp )}{J_p (zp)}
\right]\right\}.\nn\eeq In the last equation we substituted $k=zp$
in order to allow later on for a straightforward application of
the formulas for the uniform asymptotic expansion of the Bessel
functions. Again, expanding up to the order $\epsilon$ term, we
write instead \beq \eta (s;P,A ) &=& 2\epsilon \sum_p d_p(m)
p^{-s}\int\limits_C \frac {dz}{2\pi i} z^{-s} \frac
\partial
{\partial z} \nn\\
& & \left\{ - \frac 1 {z (p+1/2)} + \frac 1 {p+1/2} \frac{ J_p
'(zp ) }{J_p (zp ) } \right\} .\nn\eeq The next step in the
procedure is to shift the contour towards the imaginary axis,
turning the Bessel function $J_p$ into the Bessel function $I_p$.
In detail, we find \beq \eta (s;P,A ) &=& - \frac{ 2\epsilon} \pi
\cos \left( \frac{ \pi s} 2 \right) \sum_p d_p(m) p^{-s}
(p+1/2)^{-1}
\nn\\
& & \int\limits_0^\infty dz z^{-s} \frac d {dz} \left\{ \frac 1 z
- \frac{ I_p' (zp)}{I_p (zp )} \right\} .\nn\eeq The residues of
the eta function are completely determined by the asymptotic
behaviour of the Bessel functions, see \cite{kir01} for details.
Therefor we need to introduce some additional notation dealing
with the uniform asymptotic expansion of the Bessel function
$I_p(k)$. For $p\to\infty$ with $z=k/p$ fixed, we make use of the
uniform asymptotic expansion of the Bessel function $I_p (zp)$ and
the derivative $I'_p (zp)$. In detail, the relevant results are
\cite{abra},
\begin{eqnarray}
I_p (zp) &\sim & \frac 1 {\sqrt{2\pi p}}
\frac{e^{p\eta}}{(1+z^2)^{1/4}} \left[ 1+\sum_{l=1}^\infty
\frac{u_l (t)} { p^l} \right],\label{ex4}\\
I_p '(zp) &\sim & \frac 1 {\sqrt{2\pi p}} \frac{e^{p\eta}
(1+z^2)^{1/4} }{z} \left[ 1+\sum_{l=1}^\infty \frac{v_l (t)} {
p^l} \right],\nn\eeq where \beq t=1/\sqrt{1+z^2}\text{ and }\eta =
\sqrt{1+z^2}+\ln [z/(1+\sqrt{1+z^2})].\label{ex5}
\end{eqnarray}
Let $u_0(t)=1$. We use the recursion relationship given in
\cite{abra} to determine the polynomials $u_l (t)$ and $v_l (t)$
which appear in Equations (\ref{ex4}) and (\ref{ex5}), \beq
u_{l+1} (t) &=& \frac 1 2 t^2 (1-t^2) u_l' (t) +\frac 1 8 \int_0^t
d\tau
(1-5\tau^2) u_l (\tau),\nn \\
v_l (t) &=& u_l (t) + t (t^2-1) \left[ \frac 1 2 u_{l-1} (t) + t
u_{l-1} ' (t) \right] .\nn\eeq In particular we have $$u_1 (t) =
\frac 1 8 t - \frac 5 {24} t^3, \quad \quad v_1 (t) = -\frac 3 8 t
+ \frac 7 {24} t^3 . $$ The needed leading two contributions from
the asymptotic expansion are then given by \beq B_0 (s;P ,A) &=& -
\frac{ 2\epsilon } \pi \cos \left( \frac{ \pi s} 2 \right) \sum_p
d_p(m) p^{-s} (p+1/2)^{-1} \nn\\
& &\int\limits_0^\infty dz z^{-s} \frac d {dz} \left\{ \frac 1 z
\left( 1 - \sqrt{ 1+z^2}\right) \right\} ,\nn\\
B_{-1} (s; P,A ) &=&\frac{ 2\epsilon } \pi \cos \left( \frac{ \pi
s} 2 \right) \sum_p d_p(m) p^{-s} (p+1/2)^{-1} \nn\\
& &\int\limits_0^\infty dz z^{-s} \frac d {dz} \left\{ \frac{
\sqrt{1+z^2}} z \left( \frac 1 p
\left[ v_1 (t) - u_1 (t) \right] \right) \right\} \nn\\
&=& -\frac{ \epsilon } \pi \cos \left( \frac{ \pi s} 2 \right)
\sum_p d_p(m) p^{-s-1} (p+1/2)^{-1} \nn\\
& & \int\limits_0^\infty dz z^{-s} \frac d {dz} \frac z {1+z^2} .
\nn\eeq The integrals can be evaluated with the help of the beta
function, see \cite{grad}. Using \beq \Gamma \left( -\frac{1+s} 2
\right) = -\frac \pi {\cos \left(\frac{\pi s} 2\right) \Gamma
\left( \frac {3+s} 2 \right) }\nn\eeq the answers obtained are
\beq B_0 (s;P _A) &=& - \frac{ \epsilon } \pi \cos \left( \frac{
\pi s} 2 \right) \frac{ \Gamma \left( -\frac{s+1} 2\right) \Gamma
\left( 1 + \frac s 2 \right) } {\sqrt \pi} \sum_p d_p(m)
p^{-s} (p+1/2)^{-1}  \nn\\
&=& \epsilon \frac{\Gamma \left( 1+ \frac s 2 \right) } {\sqrt \pi
\Gamma \left( \frac{ 3+s} 2 \right) } \sum_p d_p(m)
p^{-s} (p+1/2)^{-1}  \nn\\
B_{-1} (s; P_A ) &=&-\frac{ s\epsilon } 2  \sum_p d_p(m) p^{-s-1}
(p+1/2)^{-1}. \nn\eeq The remaining summations are related to the
spectrum on the sphere. Let $d:=m-1$. We define the base
zeta-function $\zeta_{S^d}$ and the Barnes zeta-function
$\zeta_{{\mathcal B}}$ \cite{barnes},
\begin{eqnarray*}
&& \zeta _{S^d} (s) = \sum_{n=0}^\infty d_n (m) p^{-2s}\text{ and
} \zeta_{{\mathcal B}} (s,a) = \sum_{n=0}^\infty d_n (m)
(n+a)^{-s}.\end{eqnarray*} We then have the relation $$\zeta_{S^d}
(s) = \frac 1 2 d_s \zeta_{{\mathcal B}} \left( 2s, \frac m 2 -1
\right).$$ Using the Barnes zeta-function, up to terms that are
irrelevant for the present purpose because their residues are
located to the left of $s=m-3$, we find \beq B_0 (s; P,A ) &=&
\frac 1 2 d_s \epsilon \frac{\Gamma \left( 1+ \frac s 2 \right)}
{\sqrt \pi \Gamma \left( \frac {3+s} 2 \right) } \left\{ \zeta
_{{\mathcal B}} \left( s+1 , \frac m 2 -1\right) - \frac 1 2 \zeta
_{{\mathcal B}} \left( s+2 , \frac m 2 -1 \right) + ... \right\}, \nn\\
B_{-1} (s; P,A ) &=& -\frac 1 4 s \epsilon d_s \left\{\zeta
_{{\mathcal B}} \left( s+2 , \frac m 2 -1 \right) + ... \right\}.
\nn\eeq This reduces the analysis of the eta function on the ball
to the analysis of $\zeta _{{\mathcal B}}(s,a)$. To compute the
relevant residues, we first express $\zeta_{{\mathcal B}} (s,a)$
as a contour integral. Let ${\mathcal C}$ be the Hankel contour.
\beq \zeta_{{\mathcal B}} (s,a) &=& \sum_{n=0}^\infty \left(
\begin{array}{c}
   d+n-1 \\
   n
\end{array}
\right) (n+a)^{-s} = \sum_{\vec m \in \nats_0^d}
(a+m_1+...+m_d)^{-s}
\nn\\
&=& \frac{\Gamma (1-s) }{2\pi} \int_{{\mathcal C}} dt \,\,
(-t)^{s-1} \frac{e^{-at}} {(1-e^{-t})^d}.\nn \eeq The residues of
$\zeta_{{\mathcal B}} (s,a)$ are intimately connected with the
generalized Bernoulli polynomials \cite{norlund}, \beq
\frac{e^{-at} } {(1-e^{-t} )^d} = (-1)^d \sum_{n=0} ^\infty
\frac{(-t)^{n-d} } {n!} B_n^{(d)} (a) .\label{ber} \eeq We use the
residue theorem to see that \beq \mbox{Res }_{s=z}\zeta_{{\mathcal
B}} (s,a) = \frac{(-1)^{d+z} }{(z-1)! (d-z)!}
                    B_{d-z}^{(d)} (a) ,\label{barn}
\eeq for $z=1,...,d$. The needed leading poles are \beq
\mbox{Res }_{s=d}\zeta_{{\mathcal B}} (s,a) &=&  \frac 1 {(d-1)!} ,\nn\\
 \mbox{Res }_{s=d-1} \zeta_{{\mathcal B}} (s,a) &=&
  \frac{d-2a}{2 (d-2)!} .\nn \eeq
This shows \beq \mbox{Res }B_0 (d-1; P,A ) &=& \frac 1 2 d_s
\epsilon \frac{ \Gamma \left( \frac m 2 \right)} {\sqrt \pi \Gamma
\left( \frac{
m+1} 2 \right) (m-2)!} , \nn\\
\mbox{Res }B_0 (d-2; P,A ) &=& \frac 1 4 d_s \epsilon \frac{
\Gamma \left( \frac{ m-1} 2 \right) (m-3)}{\sqrt \pi \Gamma \left(
\frac m 2
\right) (m-2)!} , \nn\\
\mbox{Res }B_{-1} (d-2; P,A ) &=& -\frac 1 4 d_s \epsilon \frac{
(m-3)}{(m-2)!} ,\nn\eeq and these are all the terms contributing
to the residues of $\eta$ at $s=d-1$ and $s=d-2$. Comparing with
(\ref{cm2}) and (\ref{cm16}), after suitable rearrangements of the
$\Gamma$-function \cite{grad}, we use the doubling formula \beq
\Gamma (2x) = \frac{ 2^{2x-1}} {\sqrt \pi} \Gamma (x) \Gamma
\left( x+\frac 1 2\right),\nn\eeq and $\Gamma (x+1) = x \Gamma
(x)$, we read off
\beq c_m^2 &=& - \frac 1 4 \beta (m),\nn\\
c_m^{16} &=& \frac 1 {2(m-2)} \left( 1-\frac 1 2 \pi (m-1) \beta
(m) \right) .\nn\eeq This completes the proof of Theorem
\ref{thm-1.2}.

\section*{Acknowledgements} Research of
PG was partially supported by the MPI (Leipzig, Germany). KK
acknowledges support by the Baylor University Summer Sabbatical
Program, by the Baylor University Research Committee, and by the
MPI (Leipzig, Germany). Research of JHP was supported by Korea
Science and Engineering Foundation Grant (R05-2003-000-10884-0).

\end{document}